# A Semantic Association Page Rank Algorithm for Web Search Engines

Manuel Rojas
Oklahoma State University, CS Department
mrojas@okstate.edu

**Abstract** - The majority of Semantic Web search engines retrieve information by focusing on the use of concepts and relations restricted to the query provided by the user. By trying to guess the implicit meaning between these concepts and relations, probabilities are calculated to give the pages a score for ranking. In this study, I propose a relation-based page rank algorithm to be used as a Semantic Web search engine. Relevance is measured as the probability of finding the connections made by the user at the time of the query, as well as the information contained in the base knowledge of the Semantic Web environment. By the use of "virtual links" between the concepts in a page, which are obtained from the knowledge base, we can connect concepts and components of a page and increase the probability score for a better ranking. By creating these connections, this study also looks to eliminate the possibility of getting results equal to zero, and to provide a tie-breaker solution when two or more pages obtain the same score.

## Introduction

The Semantic Web is trying to close the gap between user demand and the need for hyperlink accessibility. This approach deals with two issues: (1) common formats for integration and combination of data drawn from diverse sources, as opposed to the original Web which mainly focused on the interchange of documents; and (2) the language for recording how the data relates to real world objects. These two features allow a person, or a machine, to start off in one database and then move through an unending set of databases, which are not connected by wires but connected by topic [7]. This information networking is based on the idea of semantic associations, where one entity (node) is connected to another entity (node) by means of a relationship (an edge).

Most search engines retrieve information accurately by exploiting key content of associations in Semantic Web resources, or relations. I propose a relation-based page rank algorithm to be used in conjunction with Semantic Web search engines which relies on information that could be extracted from user queries and the ontology for a given page. Relevance score is measured as the probability that a given resource contains those relations which existed in the user's mind at the time of query definition. The idea is to use existent relations in the ontology, named "virtual links" and apply them to a set of pages to increase the probabilities of finding the implicit relations made by the user at the time of the query.

**Related Work in Semantic Page Rank Algorithms**

The idea of exploiting ontology-based annotations for information is not new; semantic search engine would consider keyword concept associations and would return a page only if keywords (or synonyms, homonyms, etc.) are found within the page and related to associated concepts. The success is measured by the "predictability" that the user would have guessed such an association exists.

In the semantic model proposed in [1], a ranking system is created based on an estimate of the probability that keywords and/or concepts within an annotated page "A" are linked to one another in a way that is the same or similar to the one in the user's mind at the time of query definition.

The ranking strategy assumes that given a query "Q", and a page "p", it is possible to build a query subgraph $G_{Q,p}$ exploiting the information available in page annotation "A". The query subgraph is an undirected weighted graph derived from G <u>where vertices not belonging to $C_Q$ are deleted</u>. The following is a small example of a "travel" ontology and how we can derive the query subgraphs from it:

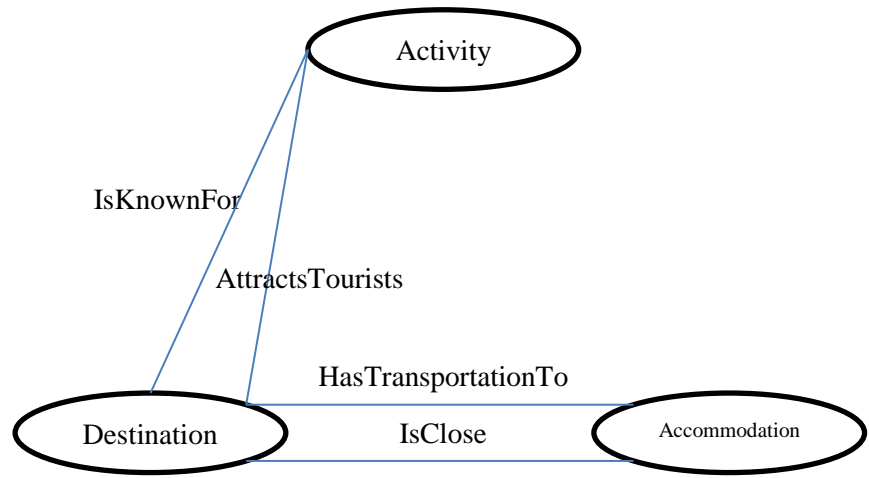

Figure 1: Travel Ontology

The undirected graph G can be defined as G(C,R), where $C_1$, $C_2$ and $C_3$ is the set of concepts that can be identified in the ontology. |C| = n is the total number of concepts available, R = {$R_{ij}$ | i = 1, …, n, j = 1, …n, j > 1} is the set of edges in the graph, and more specifically, $R_{ij} = r_{ij}^1, r_{ij}^2, … r_{ij}^m$, m < n} is the set of edges between concepts i and j. Therefore $R_{ij} = |R_{ij}|$ which are the number of relations between $C_i$ and $C_j$ in the Query subgraph $G_Q$ ($C_Q$, $R_Q$). [1] The Ontology graph G can then be represented like this:

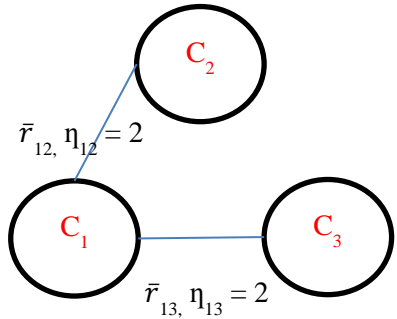

Figure 2: Ontology Graph

The ontology may have many query page subgraphs that contain the same concepts and relations in different arrangements. A weight $\delta_{ij}$ is associated to each edge to take into account the number of relations actually linking concepts i and j in the selected page (on the basis of the set of annotated relations) [1]. The concepts "destination", "activity", and "accommodation" can be related to any word in the user query. For example if the user types three terms in the search bar: "Rome", "historic-center" and "hotel", the idea is to pair each term with a concept from the ontology:

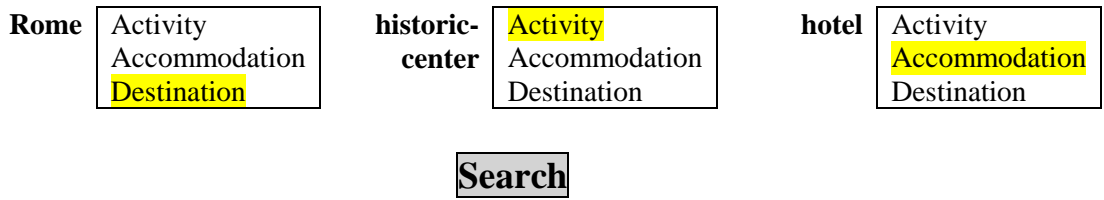

Figure 3: Search Bar

As shown above, this is done by selecting from a drop-down list. The user pairs Rome-Destination, historic center-Activity, and hotel-Accommodation. An example of two pages that come up from this configuration:

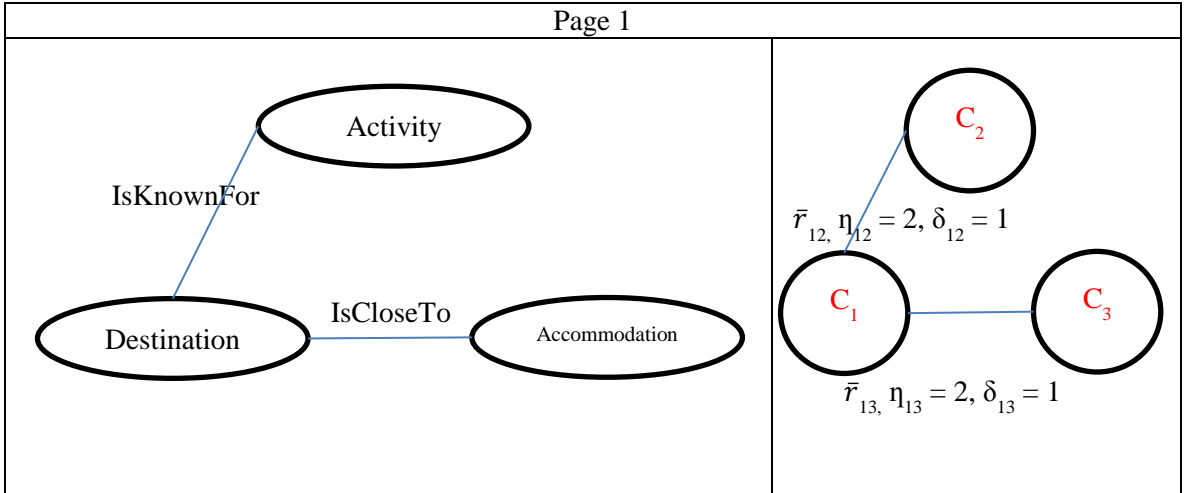

Figure 4: Page 1 extracted from ontology

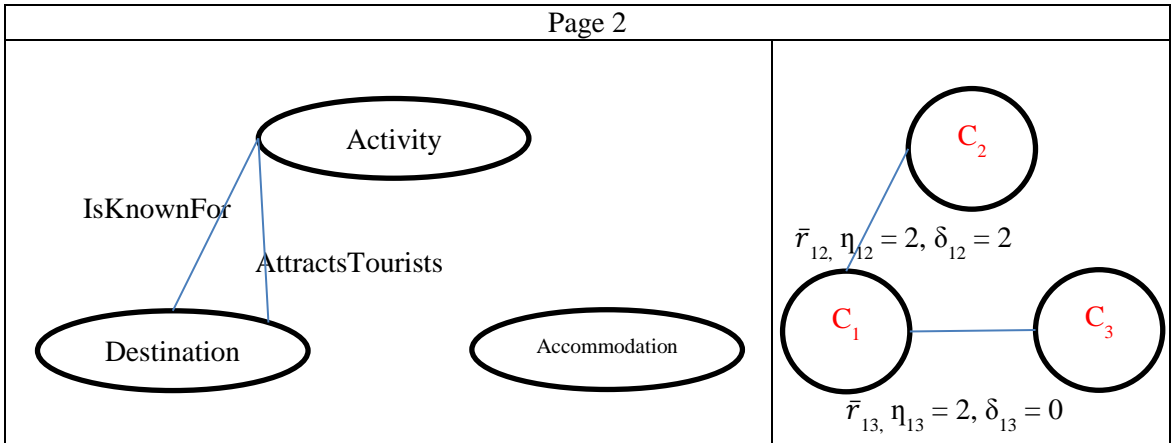

Figure 5: Page 2 extracted from ontology

Now we compute $P(\bar{r}_{ij}, Q, p)$, which is the probability of finding in a particular page p a relation $\bar{r}_{ij}$ between concepts i and j that could be the one of interest to the user (because of query Q). According to the probability theory, this can be defined as $P(\bar{r}_{ij}, p) = \delta_{ij} / \eta_{ij} = T_{ij}$. Based on the considerations above, we can compute the joint probability $(P(\bar{r}_{ij}, p) \cap P(\bar{r}_{ij}, p))$, where i and j are different for every P ( joining different edges). The dependency on the query Q is due to the fact that *only concepts given in Q are taken into account*. For the example shown in Figure 5 we could calculate the joint probabilities of the two pages as follows:

| Page 1 ($p_1$) | Page 2 ($p_2$) |
|---|---|
| Relation Probabilities: | Relation Probabilities: |
| $P(\bar{r}_{ij}, p) = \delta_{ij} / \eta_{ij} = T_{ij}$ | $P(\bar{r}_{ij}, p) = \delta_{ij} / \eta_{ij} = T_{ij}$ |
| $P(\bar{r}_{12,} p_1) = 1 / 2 = 0.5$ | $P(\bar{r}_{12,} p_2) = 2 / 2 = 1$ |
| $P(\bar{r}_{13,} p_1) = 1 / 2 = 0.5$ | $P(\bar{r}_{13,} p_2) = 0 / 2 = 0$ |
| Joint Probability: | Joint Probability: |
| $P(Q,p_1) = (P(\bar{r}_{12,} p_1)) \cap (P(\bar{r}_{13,} p_1))$ | $P(Q,p_2) = (P(\bar{r}_{12,} p_2)) \cap (P(\bar{r}_{23,} p_2)) \cap (P(\bar{r}_{13,} p_2))$ |
| $= ((1/2) \cap (1/2))$ | $= ((2/2) \cap (1/2) \cap (1/2))$ |
| $= (0.5 * 0.5)$ | $= (1 * 0)$ |
| $= 0.25$ | $= 0$ |

Table 1: Relation and Joint Probabilities

In Table 1, Page 1 would get ranked first followed by Page 2. This is a simple scenario to show how joint probability works, but since the attempt is to guess which one was the relation that the user made at the time of the query, all possibilities have to be considered. Also, *a way for assigning a score different than zero to pages in which there exists concepts not related to other concepts will have to be identified.*

If each concept is related to at least another concept in the query; this is equivalent to considering all the possible spanning forests, a collection of spanning trees of connected components in the graph, for page subgraph $G_{Q,p}$ given the query Q. We call $SF_{Q,p}^f$ the $f^{th}$ page spanning forest computed over $G_{Q,p}$. We define $P(SF_{Q,p}^f)$ as the probability that $SF_{Q,p}^f$ is the spanning forest of interest to the user. It is necessary to generate all the possible spanning forest of length "l" for each page to calculate the probability $P(Q,p,l)$ and determine the page ranking.

The method defines $SF_{Q,p}(l)$, which is the set including all the constrained spanning forests for a given number of edges l ($1 \leq l \leq C_{Q,p}$). The cardinality of this set is $\sigma_{Q,p}(l) = |SF_{Q,p}(l)|$. Finally, let $SF_{Q,p}^f(l)$ be defined as the $f^{th}$ spanning forest originated from the page subgraph for the given query Q and page p and a specific number of edges l. When current length "l" is equal to the maximum length of a spanning forest of the page subgraph, this corresponds to a page spanning forest. Otherwise, it is a constrained page spanning forest. Based on these considerations, a constrained relevance score for a page p is defined as [1]:

$$P(Q,p,l) = P \left( \bigcup_{f=1}^{SFQ,p\,(l)} \left( \cap \{\bar{r}ij, p \mid \bar{r}ij, p \in SF_{Q,p}^f(l)\} \cap SF_{Q,p}^f(l) \right) \right.$$

$$= \sum_{f=1}^{SFQ,p\,(l)} \prod_{\bar{r}ij,p \in SF_{Q,p}^f(l)} P(\bar{r}ij, p) * P\left(SF_{Q,p}^f(l)\right), \text{ where } P\left(SF_{Q,p}^f(l)\right) = 1/\sigma_{Q,p}(l)$$

This is the constrained page relevance score since its value depends on the value of length "l". By iteratively considering all the constrained spanning forests of the same length, the constraint is relaxed from having all the concepts related in some way to other concepts within the page. As soon as a value different than zero is obtained for P (Q, p, l), it can be assumed that this corresponds to a "final" relevance score for that page. However, since P (Q, p, l) is computed as a probability, this provides $0 \leq P(Q, p, l) \leq 1$. Thus, P (Q, p, l) cannot be directly used to compare one page in the result set with the remaining ones, but the information can be exploited on l to create several relevance classes unequivocally. By reducing the value of l, as soon as value different than zero for P (Q, p, l) is obtained, the page relevance score, page score, can be computed as:

$$PS_{Q,p} = P(Q, p, \max(l)) + \max(l) \mid P(Q, p, l) \neq 0$$

In this way, each relevance class contains pages with a score in the range | l, l + 1 |, pages within the same class are directly comparable, and the final result set can be ordered by decreasing values of the page score [1].

There are three key aspects previously highlighted that are worth noting, and will attempt to be improved in the new method:

1. The query subgraph is an undirected weighted graph derived from G *where vertices not belonging to $C_Q$ are deleted.*

    The new method will not delete vertices that do not belong to the query.

2. The dependency on the Query Q is due to the fact that *only concepts given in Q are taken into account.*

    The new method will attempt to include concepts that do not belong to the Query Q, but are still part of the Ontology knowledge base. The nodes will be linked by using "virtual links" obtaining spanning trees with nodes that are linked to the nodes involved in the Query.

3. *A way for assigning a score different than zero to pages in which there exists concepts not related to other concepts will have to be identified.*

Functions will be used in order to eliminate the possibility of zero results in a given page. The effects caused by edge removal will be studied in this paper, and an analytical comparison between the old and the new method will be provided to determine the best ranking strategy using a semantic association approach. The new semantic format of the Web allows building a huge repository map of relationships that can be exploited in the proposed method. As a result, better probabilities can be provided to the user by determining relationships in the user's mind at the time of the query.

**The ranking method**

To evaluate the feasibility of this new method, a controlled Semantic Web environment was constructed. To do this, we must generate controlled ontologies and page subgraphs, and then modify its relations in order to make it more suitable for demonstrating the method's functionality. The architecture workflow will look like this:

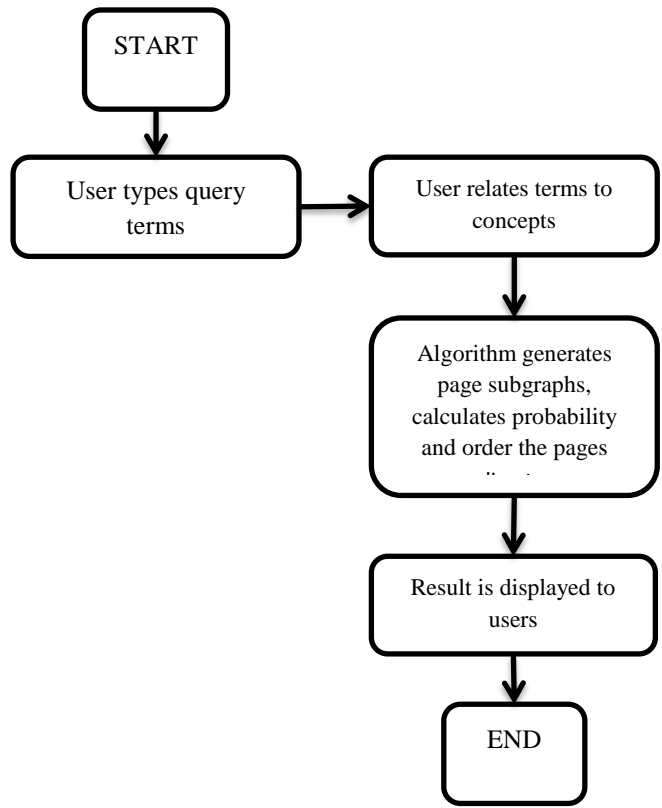

Figure 6: Architecture Workflow

The architecture is composed of a basic cycle where the user can type the query terms and then associate those terms with the concepts available in the ontology. These concepts will be limited to the ontology so that a choice from a dropdown list must be made, even if the association is not relevant to the meaning of the user. Similar to a real search, the results are limited to the resources in the Web, like the architecture proposed in the new method. Once the query terms and/or concepts are associated, the algorithm generates page subgraphs, calculates probabilities for each one, and orders them according to their score.

**Algorithm**

1.     Set variables

```
Set Nij = Number of relations of Ontology

Set Rqp = Number of relations of Page Subgraph

Set Cqp = Number of concepts of Page Subgraph

Label all edges in graph from 1 to Rqp

Set Ne = Nij // Number of relations linking concepts i and j in
Ontology

Set De = Dij // Number of relations linking concepts i and j in Page
Subgraph

// Set Te = Dij / Nij (Relation probability for edge e)
for (i = 0; i < D; i++ )

// If edge weight equals 0 then its weight will be 0.5 divided by total
number of edges

if (D[i] == 0)
  D[i] = 0.5 / Rqp
  T[i] = D[i] / N[i]

Set PS[length]; // Page Score for each Probability length
```

Table 2: New Method Algorithm – Set Variables

$R_{Q,p}$ and $C_{Q,p}$ are the total relations and number of concepts respectively on a given page subgraph. We label each $R_{ij}$ from 1 to $R_{Q,p}$, this way we can use and transverse the undirected graph in order. We then proceed to calculate the probability of each edge, which is the number of relations of the page subgraph divided by number of relations in the ontology.

2. Visit every edge to discover every spanning tree. Always start with length 1.

```
for (e = 1; e <= N; e++)
    e = visited
    visit (e, 1, Te)
    W[1] = W[1] + Te
    E[1] = E[1] + 1
    e = not visited

// Calculate Page Score PS for each spanning tree of length l
PS[l] = W[l]/E[l]
```
Table 3: New Method Algorithm – Main for-loop

The main loop visits each edge in order until the last edge is visited. Each iteration makes a call to visit, starting with length = 1. The subroutine will recursively visit all edges from the given starting point. Each time a "visit" finishes, the probability and number of spanning trees of length 1 is calculated. After all the spanning forests and probabilities have been calculated, the best option is calculated and saved as the page score.

3. Recursive subroutine visit

```
Subroutine visit (e, length, s)

a = e + 1

while (a <= Rqp and length <= Cqp - 1)
    if (a != visited and a != cycle)
      a = visited
      visit (a,length+1,sXT)
      // if the edges are all connected then
      // calculate probability for spanning tree
      if (connected_component(sXT) == 1)
        W[length+1] = W[length+1] + s
        E[length+1] = E[length+1] + 1
        a = notvisited
        a++
    else
        a++
```
Table 4: Recursive Subroutine visit

The subroutine recursively visits the next edge in order. If no more edges can be visited, or the graph has ran into a cycle, a spanning forest cannot have cycles by definition, then it will stop and calculate the probability and spanning forest of length l and save them in their respective variables W and E. Notice the if statement that checks for connected components. This step is necessary to check whether the spanning forest graph at hand is connected or not. If the elements are not connected, then by definition it is not a spanning forest. Therefore it will not be included in the page score calculations.

In order to modify the three key aspects underlined earlier "virtual links" are created between one or more concepts of a given page. These virtual links are not made up, but they are found in the ontology and can only be created if two or more pages have the same concepts. The following illustration shows how this can be done:

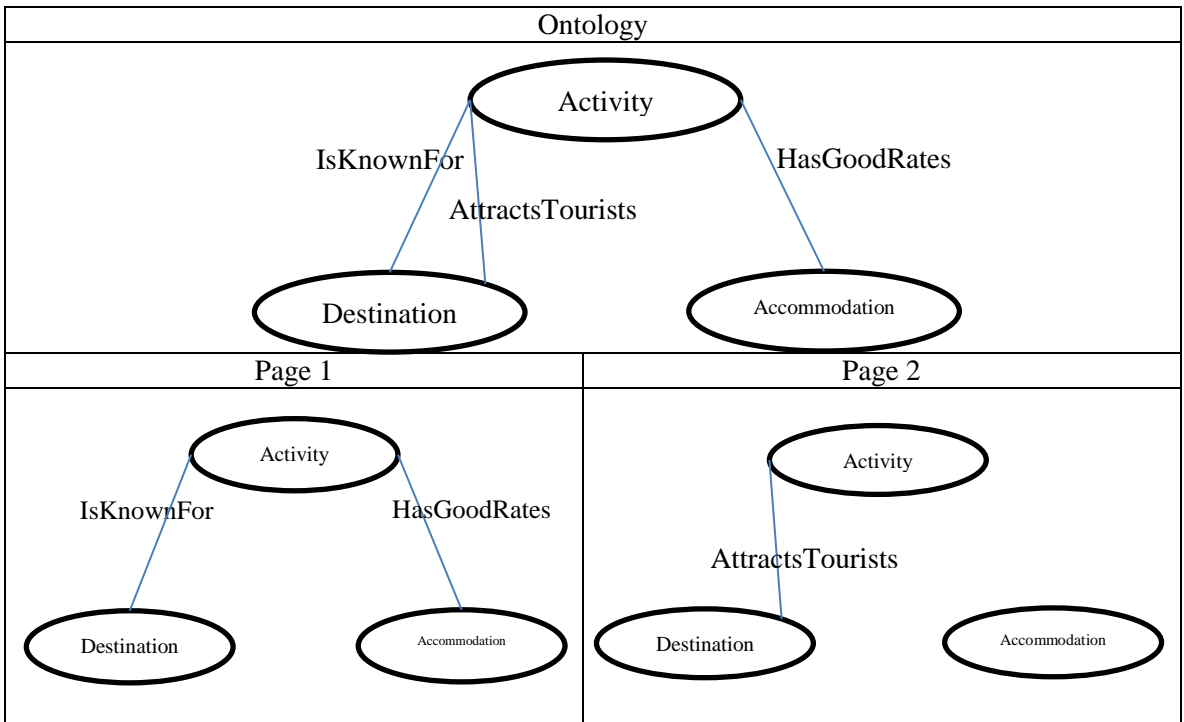

Figure 7: Example 1 - Two pages before virtual links are applied

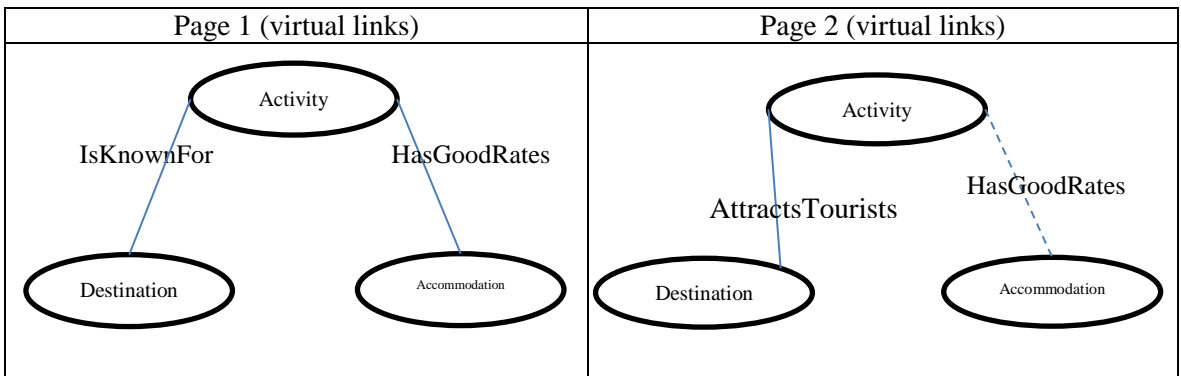

Figure 8: Example 1 - Two pages after virtual links are applied

As it shown in Figure 5, Page 1 and Page 2 have two and one relation respectively. This means that Page 2 would not have a link to the "Accommodation" node, therefore a shorter spanning forest probability $P(SF_{Q,p}^f)$ of length = 1 would be calculated for it. The idea behind the new method relies on using the information found in a given ontology to boost the probabilities of a given page. The line-dashed virtual link created in Page 2 is an example of how the "Accommodation" node can still be accessed by using a link that it is not explicitly expressed in the Page, but is found in the ontology because it is expressed in another Page. It would be easier for to define a relation between "Activity" and "Accommodation" to give more meaning to a Page, but the machine could not make this association if the information is not given to it. The virtual link value is obtained by giving a 0.5 probability to the virtual edge, and then dividing the result by the total number of edges in the ontology. From this, a page that has "true" relations would still have a better probability for a higher ranking score than a page with virtual links. Also notice that there are no virtual links between "Destination" and "Accommodation" in either one of the pages. This is because there are no links in the ontology that was used to create them. Also these virtual links are only added between concepts that do not have true links between them. Below are the probability scores for the pages shown in Figure 5:

| Page 1 | Page 2 |
|---|---|
| 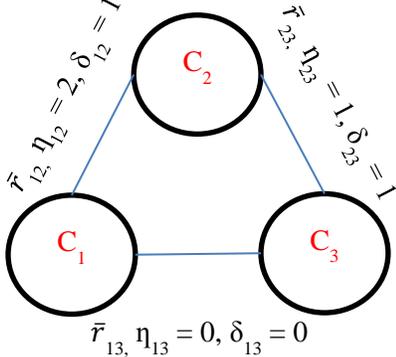 | 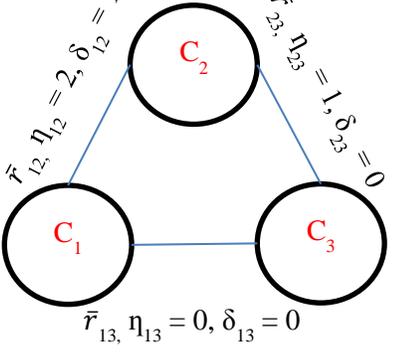 |
| $P(Q,p_1,2) = [((P(\bar{r}_{12} \cap \bar{r}_{23}) \cap P(SF_{Q,p1}^1)) \cup (P(\bar{r}_{23} \cap P(\bar{r}_{13}) \cap P(SF_{Q,p1}^2)) \cup ((P(\bar{r}_{12} \cap \bar{r}_{13}) \cap P(SF_{Q,p1}^3))] / \text{б}_{Q,p1} = [(0.5 * 1) + (1 * 0) + (0.5 * 0)] / 3$ <br> $= 0.5 / 3 = 0.16666$ <br> Add the length = 2 to the result <br> $\Rightarrow P(Q,p_1,2) = 0.166 + 2 = 2.16666$ | $P(Q,p_2,1) = [(P(\bar{r}_{12}) \cap P(SF_{Q,p3}^1)) \cup (P(\bar{r}_{23}) \cap P(SF_{Q,p3}^2)) \cup (P(\bar{r}_{13}) \cap P(SF_{Q,p3}^3))] / \text{б}_{Q,p2}$ <br> $= [(0.5 + 0 + 0)] / 3$ <br> Add the length = 2 to the result <br> $\Rightarrow P(Q,p_2,1) = 0.66666 + 1 = 1.66666$ |

Table 5: Example 2 - Two pages before virtual links are applied

| Page 1 (virtual links) | Page 2 (virtual links) |
|---|---|
| 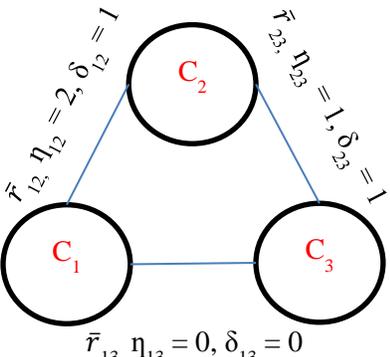 | 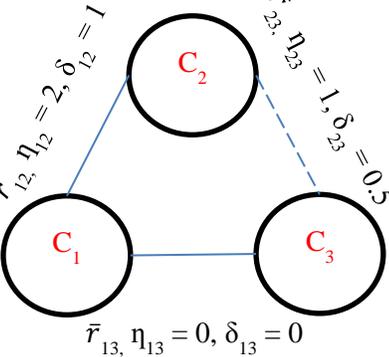 |
| $P(Q,p_1,2) = [((P(\bar{r}_{12} \cap \bar{r}_{23}) \cap P(SF^1_{Q,p1})) \cup (P(\bar{r}_{23} \cap P(\bar{r}_{13}) \cap P(SF^2_{Q,p1})) \cup ((P(\bar{r}_{12} \cap \bar{r}_{13}) \cap P(SF^3_{Q,p1}))] / \delta_{Q,p1}$ = $[(0.5 * 1) + (1 * 0) + (0.5 * 0)] / 3$ <br> $= 0.5 / 3 = 0.166$ <br> Add the length = 2 to the result <br> => $P(Q,p_1,2) = 0.16666 + 2 = 2.16666$ | $P(Q,p_2,2) = [((P(\bar{r}_{12} \cap \bar{r}_{23}) \cap P(SF^1_{Q,p2})) \cup (P(\bar{r}_{23} \cap P(\bar{r}_{13}) \cap P(SF^2_{Q,p2})) \cup ((P(\bar{r}_{12} \cap \bar{r}_{13}) \cap P(SF^3_{Q,p2}))] / \delta_{Q,p2}$ = $[(0.5 * 0.5) + (0.5 * 0) + (0.5 * 0)] / 3$ <br> $= 0.25 / 3 = 0.08333$ <br> Add the length = 2 to the result <br> => $P(Q,p_2,2) = 0.08333 + 2 = 2.08333$ |

Table 6: Example 2- Two pages after virtual links are applied

While the ranking between Page 1 and Page 2 remains the same after adding the virtual link, a slight difference in scores can be observed. The virtual link in Page 2 allows connecting concepts $C_2$ and $C_3$, thus increasing the length of the spanning forest and the page probability.

The new method allows for new "implicit" relations to form, and longer spanning forests to be created. Below is another example where the new method may be useful for two pages with similar information:

| Ontology |
|---|
| (graph with nodes $C_0, C_1, C_2, C_3, C_4, C_5$ and edges labeled: $\bar{r}_{05}, \eta_{05}=2$; $\bar{r}_{01}, \eta_{01}=2$; $\bar{r}_{04}, \eta_{04}=2$; $\bar{r}_{45}, \eta_{45}=2$; $\bar{r}_{12}, \eta_{12}=2$; $\bar{r}_{13}, \eta_{13}=2$; $\bar{r}_{23}, \eta_{23}=2$) |

| Page 1 | Page 2 |
|---|---|
| (graph with nodes $C_0, C_1, C_2, C_3$; edges: $\bar{r}_{01}, \eta_{01}=2, \delta_{01}=1$; $\bar{r}_{12}, \eta_{12}=2, \delta_{12}=1$; $\bar{r}_{13}, \eta_{13}=2, \delta_{13}=1$; $\bar{r}_{23}, \eta_{23}=2, \delta_{23}=1$) | (graph with nodes $C_0, C_1, C_2, C_3, C_4, C_5$; edges: $\bar{r}_{05}, \eta_{05}=2, \delta_{05}=1$; $\bar{r}_{04}, \eta_{04}=2, \delta_{04}=1$; $\bar{r}_{45}, \eta_{45}=2, \delta_{45}=1$; $\bar{r}_{12}, \eta_{12}=2, \delta_{12}=1$; $\bar{r}_{13}, \eta_{13}=2, \delta_{13}=1$; $\bar{r}_{23}, \eta_{23}=2, \delta_{23}=1$) |
| $P(Q,p_1,3) = ((P(\bar{r}_{01} \cap \bar{r}_{12} \cap \bar{r}_{23}) \cap P(SF^1_{Q,p2}))$ U $(P(\bar{r}_{23} \cap \bar{r}_{13} \cap \bar{r}_{13}) \cap P(SF^2_{Q,p2}))$ U $((P(\bar{r}_{13} \cap \bar{r}_{01} \cap \bar{r}_{12} \cap P(SF^3_{Q,p2}))$<br>Define $\sigma_{Q,p1}$ as the number of spanning forests for $G_{Q,p} \Rightarrow SF^1_{Q,p1} = SF^2_{Q,p1} = SF^3_{Q,p1} = 1/\sigma_{Q,p1}$ | $P(Q,p_1,2) = ((P(\bar{r}_{12} \cap \bar{r}_{23}) \cap P(SF^1_{Q,p2}))$ U $(P(\bar{r}_{23} \cap P(\bar{r}_{13}) \cap P(SF^2_{Q,p2}))$ U $((P(\bar{r}_{12} \cap \bar{r}_{13}) \cap P(SF^3_{Q,p2}))$<br>Define $\sigma_{Q,p1}$ as the number of spanning forests for $G_{Q,p} \Rightarrow SF^1_{Q,p1} = SF^2_{Q,p1} = SF^3_{Q,p1} = 1/\sigma_{Q,p1}$ |

| => [((P($\bar{r}_{01}$ ∩ $\bar{r}_{12}$ ∩ $\bar{r}_{23}$) ∩ P($SF^1_{Q,p2}$)) U (P($\bar{r}_{23}$ ∩ $\bar{r}_{13}$ ∩ $\bar{r}_{13}$) ∩ P($SF^2_{Q,p2}$)) U ((P($\bar{r}_{13}$ ∩ $\bar{r}_{01}$ ∩ $\bar{r}_{12}$ ∩ P($SF^3_{Q,p2}$)) / $\sigma_{Q,p1}$] = [(0.5 * 0.5 * 0.5) + (0.5 * 0.5 * 0.5) + (0.5 * 0.5 * 0.5)] / 3 <br> = 0.375 / 3 = 0.125 <br> Add the length = 2 to the result => P(Q,$p_1$,3) = 0.125 + 3 = 3.125 | => [((P($\bar{r}_{12}$ ∩ $\bar{r}_{23}$) ∩ P($SF^1_{Q,p1}$)) U (P($\bar{r}_{23}$ ∩ P($\bar{r}_{13}$) ∩ P($SF^2_{Q,p1}$)) U ((P($\bar{r}_{12}$ ∩ $\bar{r}_{13}$)∩ P($SF^3_{Q,p1}$))] / $\sigma_{Q,p1}$ = [(0.5 * 0.5) + (0.5 * 0.5) + (0.5 * 0.5)] / 3 <br> = 0.75 / 3 = 0.25 <br> Add the length = 2 to the result => P(Q,$p_1$,2) = 0.25 + 2 = 2.25 |
|---|---|

Table 7: Example 3 – Two pages before virtual links are applied

Table 7 shows how Page 1 has a higher spanning forest of length 3, therefore it is ranked first before page 2. Considering the number of concepts and relations for both pages, Page 2 actually has two more concepts and two more relations than Page 1. Therefore Page 2 has information than Page 1 overall. Page 2 also has two disconnected pages with spanning forest length = 3, but when the old method calculates the probabilities, it interprets both trees as a separate component, providing the maximum spanning forest length = 3. Therefore, it is feasible to imply that there is a link from $C_0$ to $C_1$ if there was relevant information on the page. Since the Ontology contains this link from Page 1, it can be used in Page 2 to create a virtual link:

| Page 1 (virtual link) | Page 2 (virtual link) |
|---|---|
| (graph with nodes $C_0, C_1, C_2, C_3$; edges $\bar{r}_{01}, \eta_{01}=2, \delta_{01}=1$; $\bar{r}_{12}, \eta_{12}=2, \delta_{12}=1$; $\bar{r}_{13}, \eta_{13}=2, \delta_{13}=1$; $\bar{r}_{23}, \eta_{23}=2, \delta_{23}=1$) | (graph with nodes $C_0, C_1, C_2, C_3, C_4, C_5$; edges $\bar{r}_{05}, \eta_{05}=2, \delta_{05}=1$; $\bar{r}_{01}, \eta_{01}=2, \delta_{01}=0.5$; $\bar{r}_{04}, \eta_{04}=2, \delta_{04}=1$; $\bar{r}_{45}, \eta_{45}=2, \delta_{45}=1$; $\bar{r}_{12}, \eta_{12}=2, \delta_{12}=1$; $\bar{r}_{13}, \eta_{13}=2, \delta_{13}=1$; $\bar{r}_{23}, \eta_{23}=2, \delta_{23}=1$) |
| P(Q,$p_1$,3) = ((P($\bar{r}_{01}$ ∩ $\bar{r}_{12}$ ∩ $\bar{r}_{23}$) ∩ P($SF^1_{Q,p2}$)) U (P($\bar{r}_{23}$ ∩ $\bar{r}_{13}$ ∩ $\bar{r}_{13}$) ∩ P($SF^2_{Q,p2}$)) U ((P($\bar{r}_{13}$ ∩ $\bar{r}_{01}$ ∩ $\bar{r}_{12}$ ∩ P($SF^3_{Q,p2}$)) | P(Q,$p_2$,5) = ((P($\bar{r}_{01}$ ∩ $\bar{r}_{04}$ ∩ $\bar{r}_{05}$ ∩ $\bar{r}_{12}$ ∩ $\bar{r}_{23}$) ∩ P($SF^1_{Q,p2}$)) U ((P($\bar{r}_{01}$ ∩ $\bar{r}_{04}$ ∩ $\bar{r}_{05}$ ∩ $\bar{r}_{12}$ ∩ $\bar{r}_{13}$) ∩ P($SF^2_{Q,p2}$)) U ((P($\bar{r}_{01}$ ∩ $\bar{r}_{04}$ ∩ $\bar{r}_{05}$ ∩ $\bar{r}_{23}$ ∩ $\bar{r}_{13}$) ∩ |
| Define $\sigma_{Q,p1}$ as the number of spanning forests for $G_{Q,p} \Rightarrow SF^1_{Q,p1} = SF^2_{Q,p1} = SF^3_{Q,p1} = 1/\sigma_{Q,p1}$ <br> $\Rightarrow [((P(\bar{r}_{01} \cap \bar{r}_{12} \cap \bar{r}_{23}) \cap P(SF^1_{Q,p2})) \cup (P(\bar{r}_{23} \cap \bar{r}_{13} \cap \bar{r}_{13}) \cap P(SF^2_{Q,p2})) \cup ((P(\bar{r}_{13} \cap \bar{r}_{01} \cap \bar{r}_{12} \cap P(SF^3_{Q,p2})) / \sigma_{Q,p1}] = [(0.5 * 0.5 * 0.5) + (0.5 * 0.5 * 0.5) + (0.5 * 0.5 * 0.5)] / 3$ <br> $= 0.375 / 3 = 0.125$ <br> Add the length = 2 to the result => $P(Q,p_1,3) = 0.125 + 3 = 3.125$ | $P(SF^3_{Q,p2})) \cup ((P(\bar{r}_{01} \cap \bar{r}_{04} \cap \bar{r}_{45} \cap \bar{r}_{12} \cap \bar{r}_{23}) \cap P(SF^4_{Q,p2})) \cup ((P(\bar{r}_{01} \cap \bar{r}_{04} \cap \bar{r}_{45} \cap \bar{r}_{12} \cap \bar{r}_{13}) \cap P(SF^5_{Q,p2})) \cup ((P(\bar{r}_{01} \cap \bar{r}_{04} \cap \bar{r}_{45} \cap \bar{r}_{23} \cap \bar{r}_{13}) \cap P(SF^6_{Q,p2})) \cup ((P(\bar{r}_{01} \cap \bar{r}_{05} \cap \bar{r}_{45} \cap \bar{r}_{12} \cap \bar{r}_{23}) \cap P(SF^7_{Q,p2})) \cup ((P(\bar{r}_{01} \cap \bar{r}_{05} \cap \bar{r}_{45} \cap \bar{r}_{12} \cap \bar{r}_{13}) \cap P(SF^8_{Q,p2})) \cup ((P(\bar{r}_{01} \cap \bar{r}_{05} \cap \bar{r}_{45} \cap \bar{r}_{23} \cap \bar{r}_{13}) \cap P(SF^9_{Q,p2}))$ <br> Define $\sigma_{Q,p1}$ as the number of spanning forests for $G_{Q,p} \Rightarrow SF^1_{Q,p1} = SF^2_{Q,p1} = SF^3_{Q,p1} = SF^4_{Q,p1} = SF^5_{Q,p1} = SF^6_{Q,p1} = SF^7_{Q,p1} = SF^8_{Q,p1} = SF^9_{Q,p1} = 1/\sigma_{Q,p2}$ <br> $\Rightarrow [((P(\bar{r}_{01} \cap \bar{r}_{04} \cap \bar{r}_{05} \cap \bar{r}_{12} \cap \bar{r}_{23}) \cap P(SF^1_{Q,p2})) \cup ((P(\bar{r}_{01} \cap \bar{r}_{04} \cap \bar{r}_{05} \cap \bar{r}_{12} \cap \bar{r}_{13}) \cap P(SF^2_{Q,p2})) \cup ((P(\bar{r}_{01} \cap \bar{r}_{04} \cap \bar{r}_{05} \cap \bar{r}_{23} \cap \bar{r}_{13}) \cap P(SF^3_{Q,p2})) \cup ((P(\bar{r}_{01} \cap \bar{r}_{04} \cap \bar{r}_{45} \cap \bar{r}_{12} \cap \bar{r}_{23}) \cap P(SF^4_{Q,p2})) \cup ((P(\bar{r}_{01} \cap \bar{r}_{04} \cap \bar{r}_{45} \cap \bar{r}_{12} \cap \bar{r}_{13}) \cap P(SF^5_{Q,p2})) \cup ((P(\bar{r}_{01} \cap \bar{r}_{04} \cap \bar{r}_{45} \cap \bar{r}_{23} \cap \bar{r}_{13}) \cap P(SF^6_{Q,p2})) \cup ((P(\bar{r}_{01} \cap \bar{r}_{05} \cap \bar{r}_{45} \cap \bar{r}_{12} \cap \bar{r}_{23}) \cap P(SF^7_{Q,p2})) \cup ((P(\bar{r}_{01} \cap \bar{r}_{05} \cap \bar{r}_{45} \cap \bar{r}_{12} \cap \bar{r}_{13}) \cap P(SF^8_{Q,p2})) \cup ((P(\bar{r}_{01} \cap \bar{r}_{05} \cap \bar{r}_{45} \cap \bar{r}_{23} \cap \bar{r}_{13}) \cap P(SF^9_{Q,p2})) / \sigma_{Q,p2}] = [ ((0.25 * 0.5 * 0.5 * 0.5 * 0.5) + (0.25 * 0.5 * 0.5 * 0.5 * 0.5) + (0.25 * 0.5 * 0.5 * 0.5 * 0.5) + (0.25 * 0.5 * 0.5 * 0.5 * 0.5) + (0.25 * 0.5 * 0.5 * 0.5 * 0.5) + (0.25 * 0.5 * 0.5 * 0.5 * 0.5) + (0.25 * 0.5 * 0.5 * 0.5 * 0.5) + (0.25 * 0.5 * 0.5 * 0.5 * 0.5) + (0.25 * 0.5 * 0.5 * 0.5 * 0.5)) / 9] = (0.140625 / 9) = 0.015625$ <br> Add the length = 5 to the result => $P(Q,p_2,5) = 0.015625 + 5 = 5.015625$ |

Table 8: Example 3 – Two pages after virtual links are applied

Table 8 demonstrates how the new method uses the virtual link to allow the page to connect its separate components and obtain nine possible spanning tree probabilities of length = 5. Since the length of the tree is longer and more edges are reached, Page 2 obtains a better score than Page 1. The new method uses the information at hand to give the page a better value according to the information found and not just the relation's probabilities. This also raises the possibility for two cases where the information, i.e. concepts and nodes, play a key role in the final score of a page. Here are some examples:

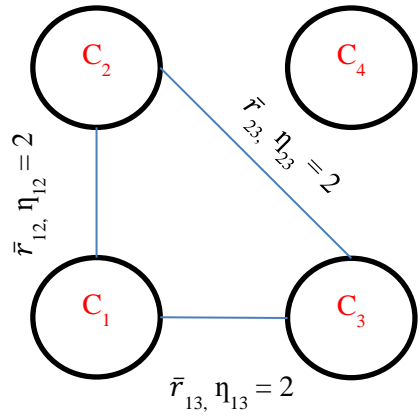

Figure 6: Example 4 - Ontology Graph

Given the Ontology on Figure 6, we can derive the following two page subgraphs:

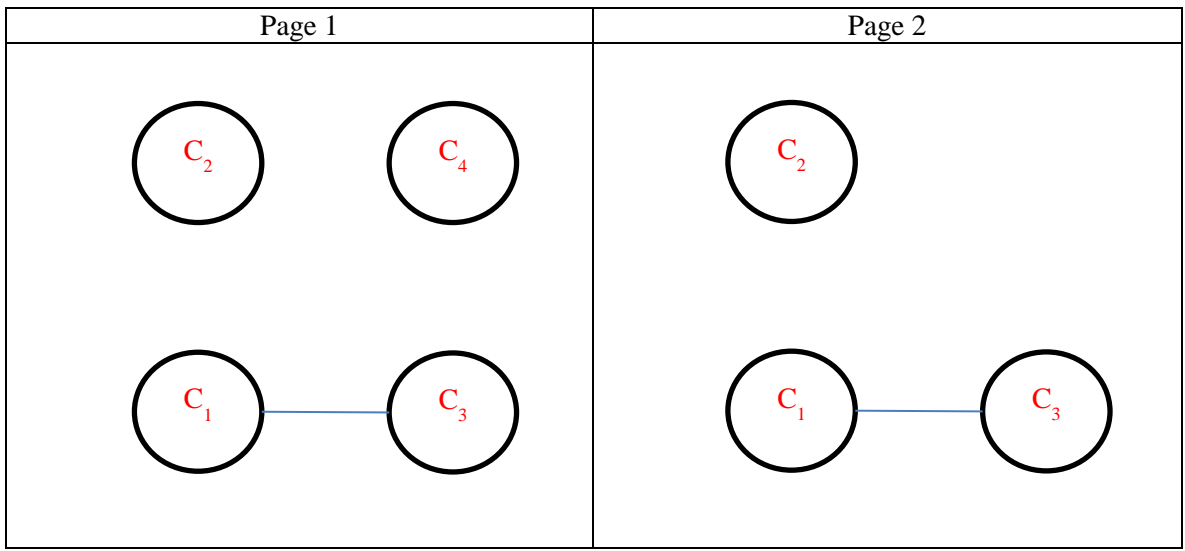

| | |
|---|---|
| (Graph: $C_2$, $C_4$, $C_1$, $C_3$ with edges labeled $\bar{r}_{12}, \eta_{12}=2, \delta_{12}=0$; $\bar{r}_{23}, \eta_{23}=2, \delta_{23}=0$; $\bar{r}_{13}, \eta_{13}=2, \delta_{13}=1$) | (Graph: $C_2$, $C_1$, $C_3$ with edges labeled $\bar{r}_{12}, \eta_{12}=2, \delta_{12}=0$; $\bar{r}_{23}, \eta_{23}=2, \delta_{23}=0$; $\bar{r}_{13}, \eta_{13}=2, \delta_{13}=1$) |
| $P(Q,p_1,2) = (P(\bar{r}12) \cap P(SF^1_{Q,p1})) \cup (P(\bar{r}23) \cap P(SF^2_{Q,p1})) \cup (P(\bar{r}13) \cap P(SF^3_{Q,p1}))$<br>Define $\sigma_{Q,p1}$ as the number of spanning forests for $G_{Q,p} \Rightarrow SF^1_{Q,p1} = SF^2_{Q,p1} = SF^3_{Q,p1} = 1/\sigma_{Q,p1}$<br>$\Rightarrow [(P(\bar{r}12) \cap SF^1_{Q,p1}) \cup (P(\bar{r}23) \cap SF^2_{Q,p1}) \cup (P(\bar{r}13) \cap SF^3_{Q,p1})] / \sigma_{Q,p1} =$<br>$= (0 + 0 + 0) / 3 = 0$<br>Add spanning tree length:<br>$= 0 + 2 = 2$ | $P(Q,p_2,2) = (P(\bar{r}12) \cap P(SF^1_{Q,p2})) \cup (P(\bar{r}23) \cap P(SF^2_{Q,p2})) \cup (P(\bar{r}13) \cap P(SF^3_{Q,p2}))$<br>Define $\sigma_{Q,p1}$ as the number of spanning forests for $G_{Q,p} \Rightarrow SF^1_{Q,p2} = SF^2_{Q,p2} = SF^3_{Q,p2} = 1/\sigma_{Q,p2}$<br>$\Rightarrow [(P(\bar{r}12) \cap SF^1_{Q,p1}) \cup (P(\bar{r}23) \cap SF^2_{Q,p2}) \cup (P(\bar{r}13) \cap SF^3_{Q,p2})] / \sigma_{Q,p2} =$<br>$= (0 + 0 + 1) / 3 = 0$<br>Add spanning tree length:<br>$= 0 + 2 = 2$ |

Table 9: Example 4 – Page Subgraphs

In this case, Page 1 and Page 2 would tie for first place since they have the same page score of 2. While the main idea is to calculate the probability of the relations formed in a user's head at the time of the query, the new method aims to take into consideration the information at hand to provide the best possible ranking. In this example, we could use the extra concept $C_4$ in Page 1 to modify the overall score of the page and provide further relevance for the values at hand. In order to do this, divide the total number of nodes found in a given page by the total number of nodes found in the ontology. This value will then be added to the probability score to produce the final page score.

This is another example when the number of relations in a given page subgraph may be used to define the final score. Consider the following ontology:

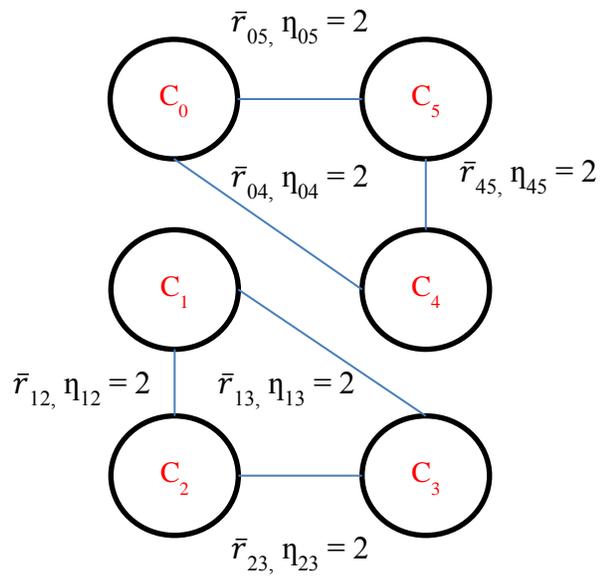

Figure 9: Example 5 - Ontology Graph

From the Ontology in Figure 17, we get the following page subgraphs:

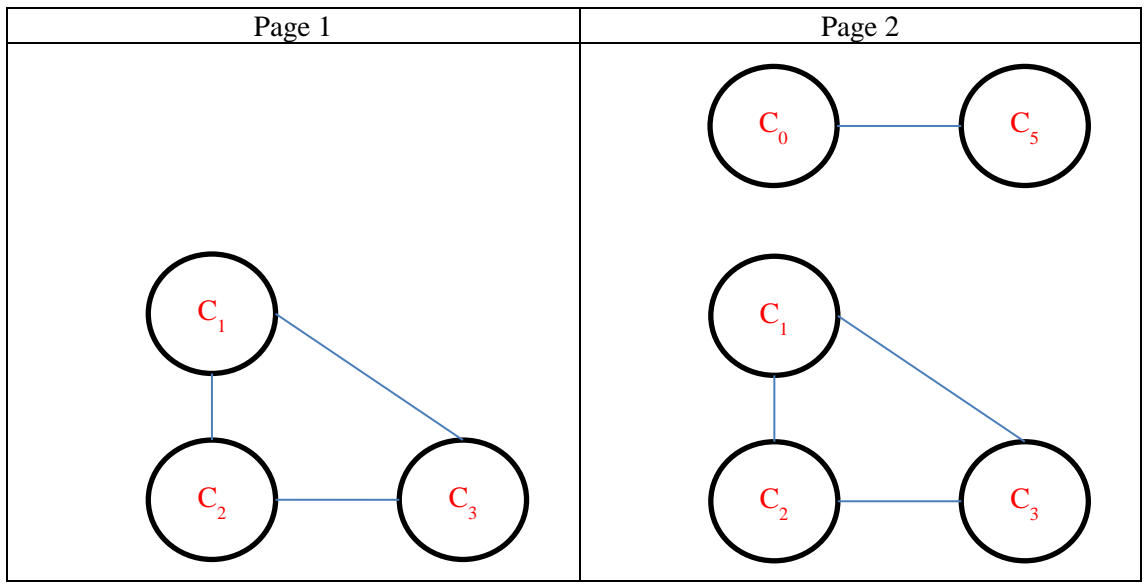

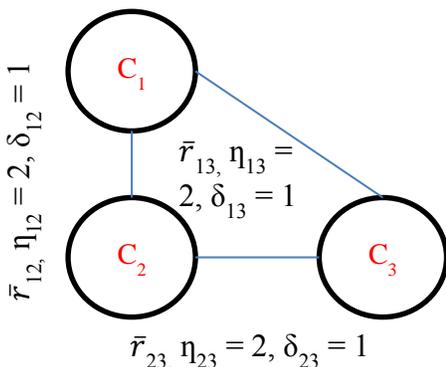

| | |
|---|---|
| $P(Q,p_1,2) = (P(\bar{r}12) \cap P(SF^1_{Q,p1})) \cup (P(\bar{r}23) \cap P(SF^2_{Q,p1})) \cup (P(\bar{r}13) \cap P(SF^3_{Q,p1}))$ <br> Define $\sigma_{Q,p1}$ as the number of spanning forests for $G_{Q,p} \Rightarrow SF^1_{Q,p1} = SF^2_{Q,p1} = SF^3_{Q,p1} = 1/\sigma_{Q,p1}$ <br> $\Rightarrow [(P(\bar{r}_{12}) \cap SF^1_{Q,p1}) \cup (P(\bar{r}_{23}) \cap SF^2_{Q,p1}) \cup (P(\bar{r}13) \cap SF^3_{Q,p1})] / \sigma_{Q,p1} =$ <br> $= (0.25 + 0.25 + 0.25) / 3$ <br> Add spanning tree length: <br> $= 0.25 + 2 = 2.25$ | $P(Q,p_2,2) = (P(\bar{r}12) \cap P(SF^1_{Q,p2})) \cup (P(\bar{r}23) \cap P(SF^2_{Q,p2})) \cup (P(\bar{r}13) \cap P(SF^3_{Q,p2}))$ <br> Define $\sigma_{Q,p1}$ as the number of spanning forests for $G_{Q,p} \Rightarrow SF^1_{Q,p2} = SF^2_{Q,p2} = SF^3_{Q,p2} = 1/\sigma_{Q,p2}$ <br> $\Rightarrow [(P(\bar{r}_{12}) \cap SF^1_{Q,p2}) \cup (P(\bar{r}_{23}) \cap SF^2_{Q,p2}) \cup (P(\bar{r}13) \cap SF^3_{Q,p2})] / \sigma_{Q,p2} =$ <br> $= (0.25 + 0.25 + 0.25) / 3$ <br> Add spanning tree length: <br> $= 0.25 + 2 = 2.25$ |

Table 10: Example 5: Page Subgraphs

As it is in the same way with the case of total number of nodes, Table 10 shows two pages that obtain the same score. By taking advantage of the information on each page, it can be determined that Page 2 has one more relation $\bar{r}_{05}$ that may be useful for the user, which Page 1 does not have. To obtain a ranking value that can be used to compare the pages, divide the total number of edges in a page subgraph by the total number of edges in the ontology. This value will then be added to the probability score to produce the final page score.

The number of edges and number of nodes function provide a way for the pages to be ranked in terms of their concepts and relations, independently from the probability score. This completely eliminates the possibility of obtaining a zero score as long as the page has at least one concept. Both of these values obtained from the functions are then combined to give the new method's *combined* final score. Since we have to take in consideration both functions (concepts and relations), the combined function allows the page to extract this information and add it to the final score value. The following table shows the comparison between the old method and new the method, with nodes, edges, and combined functions:

| 10 Pages | | | | | | | | | | |
|---|---|---|---|---|---|---|---|---|---|---|
| **Onto** | **$P_1$** | **$P_2$** | **$P_3$** | **$P_4$** | **$P_5$** | **$P_6$** | **$P_7$** | **$P_8$** | **$P_9$** | **$P_{10}$** |
| 2,0,5 | 2,0,2 | 5,2,2 | 5,2,2 | 0,3,0 | 0,2,1 | 5,1,3 | 1,5,4 | 4,0,4 | 1,3,1 | 0,5,3 |
| 4,1,5 | 4,1,2 | 0,3,3 | 5,1,0 | 1,5,4 | 0,1,2 | 1,4,3 | 2,0,3 | 0,2,2 | 0,1,3 | 5,1,1 |
| 0,5,5 | 0,5,3 | 1,0,2 | 3,0,3 | 3,2,4 | 3,1,3 | 2,3,3 | 3,0,1 | 0,1,1 | 0,4,3 | 1,0,3 |
| 0,3,5 | 0,3,1 | 4,3,4 | 0,2,3 | 2,5,0 | 2,1,3 | 1,2,0 | 4,5,0 | 5,1,3 | 5,0,2 | 2,1,3 |
| 4,5,5 | 4,5,4 | 2,3,0 | 5,0,3 | 1,3,2 | 1,5,2 | 0,3,4 | 3,5,4 | 2,1,2 | 5,2,1 | 1,3,3 |
| 5,2,5 | | | | | | | | | | |
| 1,0,5 | | | | | | | | | | |
| 4,3,5 | | | | | | | | | | |
| 2,3,5 | | | | | | | | | | |
| 5,1,5 | | | | | | | | | | |
| 1,3,5 | | | | | | | | | | |
| 2,1,5 | | | | | | | | | | |
| 3,5,5 | | | | | | | | | | |
| 4,0,5 | | | | | | | | | | |

| Old Method | New Method (Nodes) | New Method (Edges) | New Method (Combined) |
|---|---|---|---|
| PS[1] = 4.01536 | PS[6] = 5.01728 | PS[6] = 4.37442 | PS[6] = 5.37442 |
| PS[9] = 4.00576 | PS[1] = 5.01536 | PS[1] = 4.3725 | PS[1] = 5.3725 |
| PS[10] = 3.072 | PS[2] = 5.00768 | PS[2] = 4.36482 | PS[2] = 5.36482 |
| PS[8] = 3.0512 | PS[7] = 5.00768 | PS[7] = 4.36482 | PS[7] = 5.36482 |
| PS[5] = 3.0352 | PS[9] = 5.00576 | PS[9] = 4.3629 | PS[9] = 5.3629 |
| PS[7] = 3.0256 | PS[10] = 3.90533 | PS[10] = 3.42914 | PS[10] = 4.26248 |
| PS[3] = 2.072 | PS[8] = 3.88453 | PS[8] = 3.40834 | PS[8] = 4.24168 |
| PS[2] = 2.0384 | PS[5] = 3.86853 | PS[5] = 3.39234 | PS[5] = 4.22568 |
| PS[4] = 2.03657 | PS[3] = 3.85013 | PS[3] = 3.37394 | PS[3] = 4.20728 |
| PS[6] = 1.168 | PS[4] = 3.84293 | PS[4] = 3.36674 | PS[4] = 4.20008 |

Table 11: Result's Comparison

Table 11 shows the differences in ranking between the old method and the new method using the nodes and edge functions, and the final combined function. The new method ranks Page 6 ($P_6$) first, while the old method ranks $P_6$ last. Since the relation $\bar{r}_{12}$ is considered as a virtual link in the new method, this allows $P_6$ to have a longer spanning forest. If we consider the amount of information that both pages have, we can see that $P_6$ has a total of 13 relations, the most compared to any other page. The reason $P_6$ gets ranked last in the old method is because of the weight $\delta_{12} = 0$. This makes two constrained spanning forests of length = 2 and length = 3, and treats each one as a separate component. For that reason, even when $P_6$ has the most relations in the set of pages, it will be ranked last.

The new method gives a virtual value to $\delta_{12} = 0.5$. This enables the algorithm to take into consideration the edge and obtain a longer spanning forest with a better probability. Therefore, the new method ranks $P_6$ first. We can also see how $P_1$, which has a total of 12 relations and originally ranked first in the old method, still gets second in the new method. Below is another example from the same set:

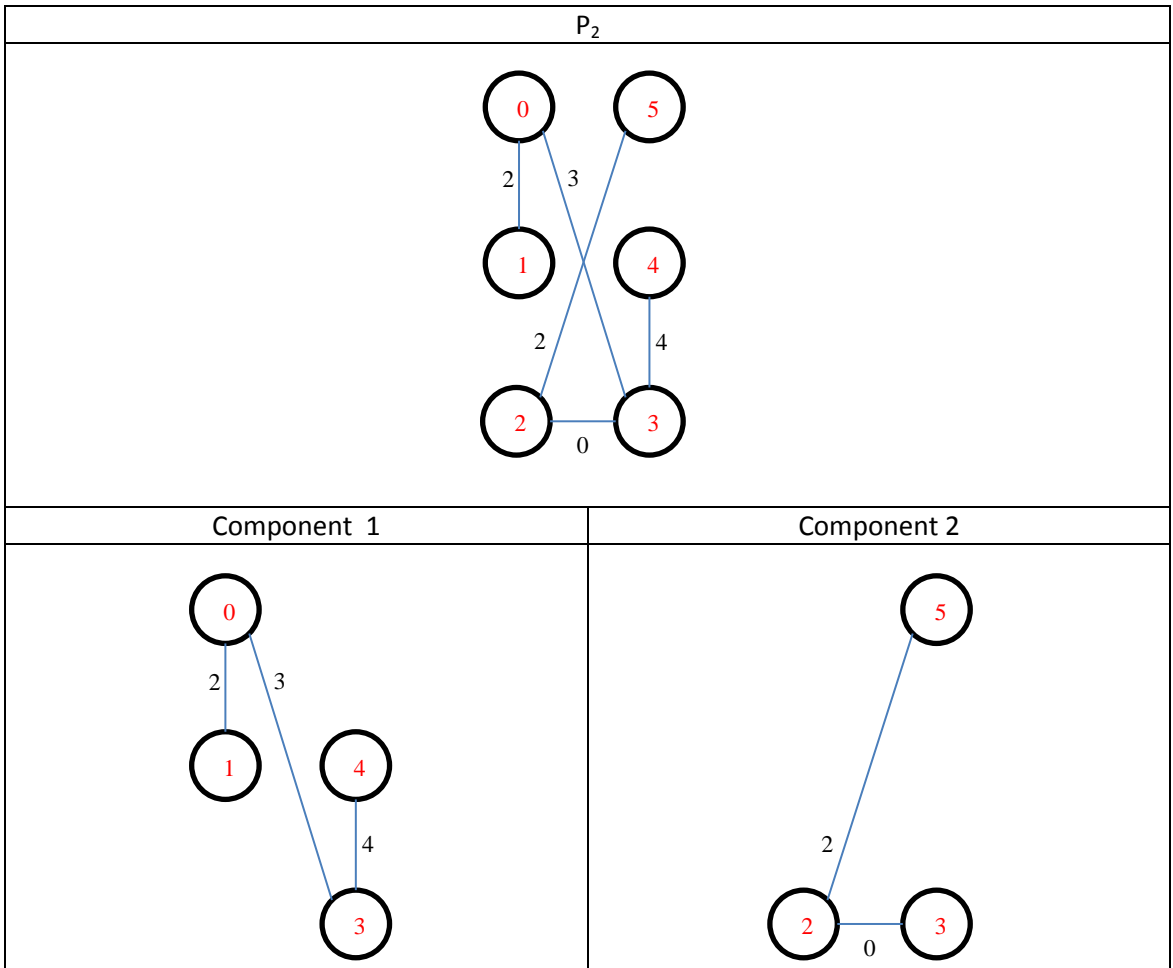

Table 12: Page 2 constrained components

Table 12 shows Page 2 ($P_2$) which contains a total of eleven relations or only two less than $P_6$. This is a prime example of how a virtual link can join two separate components ($P(\bar{r}_{01}) \cap P(SF^1_{Q,p2})$) U ($P(\bar{r}_{03}) \cap P(SF^2_{Q,p2})$) U ($P(\bar{r}_{34}) \cap P(SF^3_{Q,p2})$) and ($P(\bar{r}_{23}) \cap P(SF^1_{Q,p2})$). It converts the former into a spanning forest of length = 5 and boosts the result, allowing $P_2$ to be ranked 3$^{rd}$ overall.

Improving a probability score is determined by the size of the separate components in the new method. If a virtual link allows connecting multiple concept and relation's components to one another, then a boost in the result can be expected. If on the other hand a virtual link only connects one component to a single concept, or a smaller concept and relation's component, then page score boost will be less significant. This is true for pages $P_1$, $P_2$ and $P_6$ of the previous set of pages. The algorithm uses the virtual links and the functions to produce the best score for each page.

**Back-link method**

The two functions provide a different ranking scheme based on the total number of nodes and total number of edges. The information contained on each page along with the ontology, allots to create two different ranking schemes. Combining these two ranking schemes by adding the explicit value of nodes and edges to the page score determined in the new method, allows the final page score for each page to be obtained. It is worth noting, that while the results can be exploited to obtain a result different than zero within a page score and improve the overall ranking by using the information available in the ontology, or virtual links, in both functions the total number of nodes and total number of edges yield an equivalent ranking scheme. A comparison can then be made between these two functions to provide more than one ranking scheme. Since it is difficult to justify with a 100% certainty which relation is chosen as the relation made in the user's head at the time of the query, it cannot be assumed that one ranking system is better than the other. A feasible solution is to compare the new method with the old one method, along with a third ranking scheme. The third ranking scheme is based on back-links and it is used by search engines such as Google.

According to [9], a back-link is a "link in one direction implied from the existence of an explicit link in the other direction". A back-link may be non-directional, one-directional or bi-directional.

A search engine like Google, judges the importance of a page based on the importance of its back-links. In this simple principle resides much of Google's success. Instead of using news, expert advice, or other human communication intervention, Google lets the Web do the ranking using the Web's own linking structure. If the page $P_1$ has a total of for example 10 links to other pages, among them the page Q, then $P_1$ will transfer one tenth of its importance to Q [10].

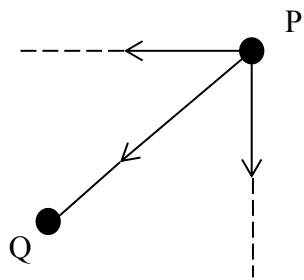

Figure 10: Backlink One-Directional Graph

Suppose that the pages that link to page Q are $P_1$, $P_2$, etc, up to page $P_n$ These pages are called the back-links of Q. Write $I_1$, $I_2$, and so forth, for the importance of the pages $P_1$, $P_2$, etc. Similarly, write $l_1$, $l_2$, and so forth, for the total number of links on page $P_1$, $P_2$, etc. Afterwards, the importance of page Q can be determined as: $I_Q = \frac{I_1}{l_1} + \frac{I_2}{l_2} + \cdots + \frac{I_n}{l_n}$. As a result, the importance of the page Q is a weighted sum of the importance of its back-links. The pages can be labeled in the Web as $P_1$, $P_2$, $P_3$, and so forth, to build an array H of numbers in which the entry corresponding to the $i^{th}$ row and $j^{th}$ column is $\frac{1}{l_j}$.

This is an example of a small web composed of four pages:

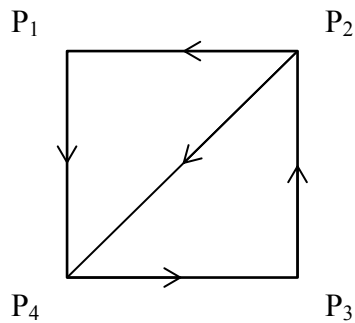

Figure 11: Small Web Composed of Four Pages

Figure 11 has the array:

$$H = \begin{pmatrix} 0 & 1/2 & 0 & 0 \\ 0 & 0 & 1 & 0 \\ 0 & 0 & 0 & 1 \\ 1 & 1/2 & 0 & 0 \end{pmatrix}$$

In the above example of matrix $H$, the second entry in the first row is 1/2 because $P_2$ links to $P_1$ and has a total of two links. If the importance of each page in known, the information can be collected and composed in a vector $I = (I_1, I_2, \ldots, I_n)$ of length n, and linear algebra stipulates that the product of a matrix $H$ and the importance vector $I$ is equal to $I$ itself [10]:

$$HI = I$$

Therefore is it possible find vector $I$ when $H$ is known and $I$ is not. The property that $HI = I$ is mathematically described by suggesting the importance vector is an eigenvector of a matrix with eigenvalue 1. In general, a vector $V$ is an eigenvector of $H$ with eigenvalue $k$, if $H$ multiplied by $H$ produces the vector $V$ with each entry multiplied by $k$ [10].

The information provided by the ontology does not directly give the back-links from one page to another. This approach infers the links by seeing which pages have the same relations. If two pages share relations then a back-link will be assigned. If more than one relation exists then more back-links will be assigned. To achieve this, relations in the ontology will have to be identified with a unique identification in order to recognize which relations are being shared by two or more pages. Since relations in the ontology and page subgraph are non-directed, then the back-links from one page to another will be non-directed as well. This means that if a back-link from $P_1$ to $P_2$ exists, then a back-link from $P_2$ to $P_1$ also exists. Here is an example of how we can provide an alternative ranking scheme for comparison:

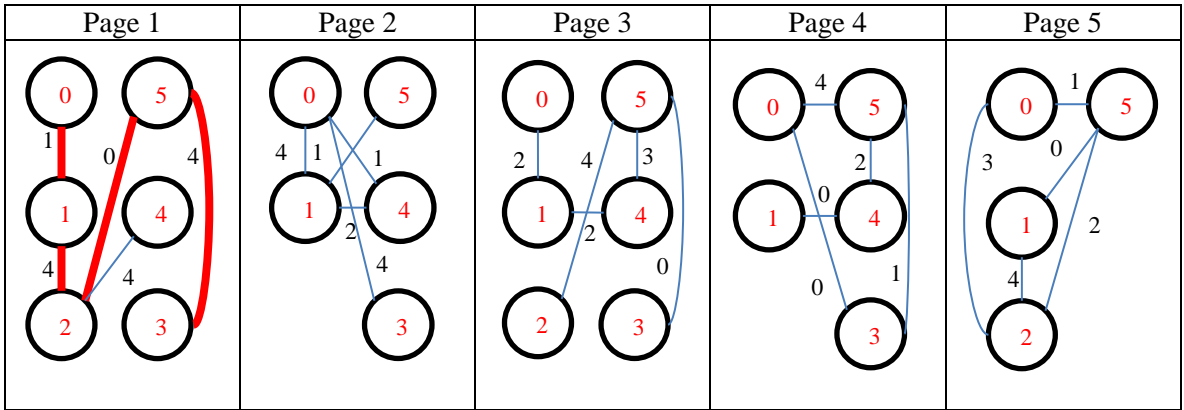

Table 12: Backlink Method: Example 1

Table 12 shows the five pages in Example 1. The highlighted edges shown in page 1 are the common links or back-links that associate Page 1 with the rest of the pages. As shown, relation $r_{24}$ is not found in any other page. Therefore it does not have a back-link and it is not highlighted in red. The back-links can be represented with matrices for each page in the following manner:

$$H_1 = \begin{pmatrix} 0 & 1 & 1 & 1 & 1/2 \\ 1 & 0 & 1/3 & 0 & 0 \\ 1 & 1/3 & 0 & 1/2 & 1/2 \\ 1 & 0 & 1/2 & 0 & 1 \\ 1/2 & 0 & 1/2 & 1 & 0 \end{pmatrix}$$

This matrix produces the following eigenvalue and eigenvector:

$\lambda_1 = 2.48113$

$V_1 = (1.43173, 0.727415, 1.11926, 1.20564, 1)$

So according to the eigenvector the pages would be ranked in the following manner:

PS[1] = 1.43173

PS[4] = 1.20564

PS[3] = 1.11926

PS[5] = 1

PS[2] = 0.727415

If same approach is applied to Examples 2 and 3, this is the following result:

**Example 2:**

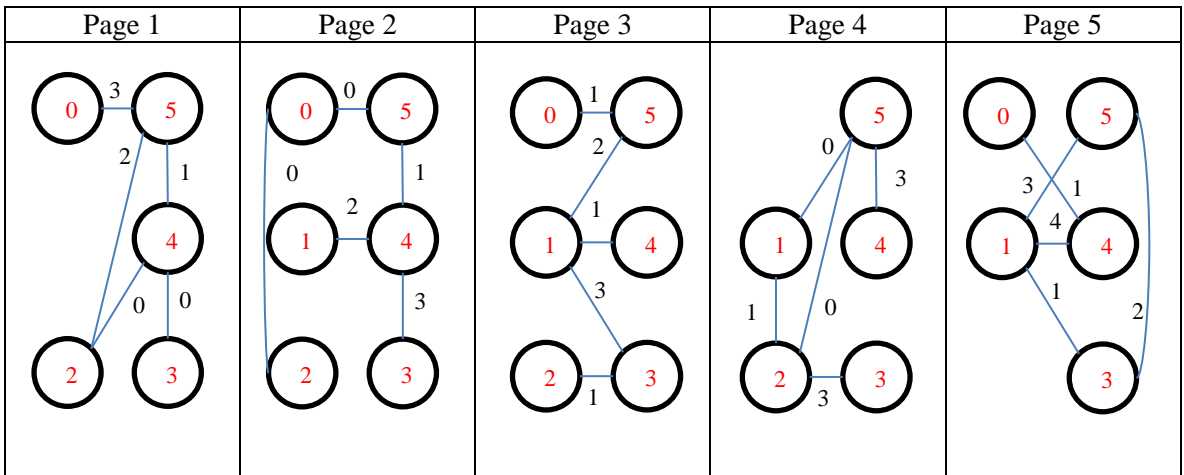

Table 13: Backlink Method: Example 2

$$H_2 = \begin{pmatrix} 0 & 1 & 1 & 1 & 0 \\ 1 & 0 & 1 & 1 & 1/2 \\ 1 & 1 & 0 & 1 & 1/4 \\ 1 & 1 & 1 & 0 & 0 \\ 0 & 1/2 & 1/4 & 0 & 0 \end{pmatrix}$$

$\lambda_1 = 3.04681$

$V_1 = (3.95945, 4.083, 4.02123, 3.95945, 1)$

PS[2] = 4.083

PS[3] = 4.02123

PS[1] = 3.95945

PS[4] = 3.95945

PS[5] = 1

**Example 3:**

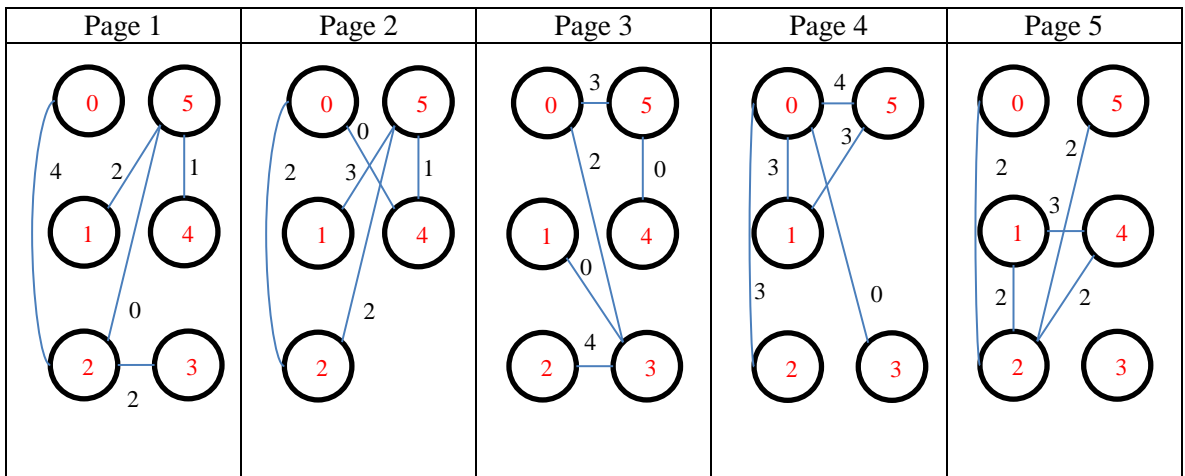

Table 24: Backlink Method: Example 3

$$H_3 = \begin{pmatrix} 0 & 1/4 & 1/2 & 1/5 & 1/2 \\ 1/4 & 0 & 0 & 1/5 & 1/6 \\ 1/2 & 0 & 0 & 1/3 & 0 \\ 1/5 & 1/5 & 1/3 & 0 & 1/2 \\ 1/2 & 1/6 & 0 & 1/2 & 0 \end{pmatrix}$$

$\lambda_1$ = 1.12237

$V_1$ = (1.08779, 0.563484, 0.772415, 0.96913, 1)

PS[1] = 1.08779

PS[5] = 1

PS[4] = 0.96913

PS[3] = 0.772415

PS[2] = 0.563484

**Ranking methods comparison**

| Ranking Methods Comparison | | | | |
|---|---|---|---|---|
| Old Method | New Method (Nodes) | New Method (Edges) | New Method (Combined) | Eigen-vector method |
| Example1 | | | | |
| PS[2] = 4.02987 | PS[3] = 6.00768 | PS[3] = 5.42435 | PS[3] = 6.42435 | PS[1] = 1.43173 |
| PS[3] = 4.0256 | PS[1] = 6.00256 | PS[1] = 5.41923 | PS[1] = 6.41923 | PS[4] = 1.20564 |
| PS[5] = 3.044 | PS[2] = 4.8632 | PS[2] = 4.44653 | PS[2] = 5.27987 | PS[3] = 1.11926 |
| PS[4] = 3.0128 | PS[4] = 4.8368 | PS[4] = 4.42013 | PS[4] = 5.25347 | PS[5] = 1 |
| PS[1] = 3.0064 | PS[5] = 3.72417 | PS[5] = 3.47417 | PS[5] = 4.14083 | PS[2] = 0.727415 |
| Example 2 | | | | |
| PS[3] = 5.00192 | PS[3] = 6.00192 | PS[3] = 5.38654 | PS[3] = 6.38654 | PS[2] = 4.083 |
| PS[5] = 4.02347 | PS[2] = 6.00048 | PS[2] = 5.3851 | PS[2] = 6.3851 | PS[3] = 4.02123 |
| PS[2] = 3.012 | PS[5] = 4.8568 | PS[5] = 4.40808 | PS[5] = 5.24142 | PS[1] = 3.95945 |
| PS[1] = 3.00686 | PS[4] = 4.83933 | PS[4] = 4.39062 | PS[4] = 5.22395 | PS[4] = 3.95945 |
| PS[4] = 2.01714 | PS[1] = 4.83613 | PS[1] = 4.38742 | PS[1] = 5.22075 | PS[5] = 1 |
| Example 3 | | | | |
| PS[5] = 4.03413 | PS[1] = 6.00256 | PS[1] = 5.38718 | PS[1] = 6.38718 | PS[1] = 1.08779 |
| PS[4] = 3.099 | PS[3] = 6.00192 | PS[3] = 5.38654 | PS[3] = 6.38654 | PS[5] = 1 |
| PS[2] = 3.0832 | PS[5] = 4.86747 | PS[5] = 4.41875 | PS[5] = 5.25208 | PS[4] = 0.96913 |
| PS[3] = 3.048 | PS[4] = 4.85973 | PS[4] = 4.41102 | PS[4] = 5.24435 | PS[3] = 0.772415 |
| PS[1] = 2.06667 | PS[2] = 4.8472 | PS[2] = 4.39848 | PS[2] = 5.23182 | PS[2] = 0.563484 |

Table 14: Ranking Methods Comparison

**Conclusion**

Search engines previously focused on the presence of keywords and statistical algorithms applied after ranking to display the results. The need for result refinement and feedback can be time consuming, and the click-expensive method causes more processing time. The Semantic Web structure allows machines to make meaningful relationships between concepts in a given webpage. This proposed concept-relations architecture provides base knowledge that can be efficiently used to retrieve specific webpages that will meet users' needs.

Different methods have been tested to exploit ontology-based annotations for information retrieval [13][14], but there is still a need for a more customizable and flexible ranking scheme to support semantic association's results. While other approaches [1][15][16] have already been taken into consideration, the work done in [1] was used as the main model for improvement and comparison. Specifically, two key aspects were attempted to be improved from this method:

1. The dependency on the Query Q is due to the fact that only concepts given in Q are taken into account. The new proposed method includes concepts that do not belong to the Query Q, but are still part of the ontology knowledge base. The nodes were linked using "virtual links", by obtaining spanning forests with nodes that were linked to the nodes involved in the query. By linking more nodes, longer spanning forests are obtained, automatically increasing the probability for a given page.
2. A way of assigning a score different than zero to pages in which there exists concepts not related to other concepts will have to be identified. The idea presented here is to provide a lesser value than a true relation ($\delta_{ij} = 1$), but a value significant enough to improve the probabilities of forming the best possible spanning forest with the information at hand. The value $\delta_{ij} = 0.5$ was the default value assigned, and then divided by the total number of relations in a given page.

To test and compare the two methods a C++ Simulator was built. Both the former and new method's algorithms were built separately, and ran against the same ontologies and set of pages. The main idea was to use the information at hand to provide a better ranking result for users. Besides taking into consideration the probability score given by the spanning forests, two different functions were added to the score to improve the results: number of nodes and number of edges. Apart from these two functions, a final combined method was added to the probability score to provide a final page score result.

Different sets of pages with small, medium, and large loads (10, 50, 100 pages respectively) were tested. The results were similar no matter the amount of pages ranked, but a few patterns were identified:

1. A longer spanning forest typically yields better results, despite the total number of nodes or edges taken into consideration. This is because the length of the forest itself is added to the score, therefore the longer the length, the greater the score.
2. The total number of nodes' value added to the probability score provides a better ranking when the pages have the same total number of relations. When two pages have the same amount of relations, they will obtain the same probability. As a result, the function will provide a tie breaker that allows the better page to be ranked.
3. The total number of edges' value added to the probability score provides a better ranking when components of a given page are not connected. These components are computed separately, by taking into account the longest spanning tree, and omitting the components that are separated from it.
4. The virtual links allows pages that have a relations of weight $\delta_{i,j} = 0$, to obtain a ranking value different than zero.
5. The new method performs better when a virtual link allows the connection between more than one edge and relation.
6. Pages that have no relations can also be evaluated in terms of their concepts, and also obtain a value different than zero.

While the new method proves to work efficiently by eliminating the possibility for zero scores, and provides a tie-breaker solution for pages that originally obtained the same result, there are also constraints observed in the new method:

1. The total amount of virtual links plays an important role in the ranking. If a page contains too many virtual links, then such page may obtain a better score than a page containing true relations and fewer virtual links.
2. The possibility of a tie between two or more pages still remains if they contain equivalent information, and no difference can be extracted from the total number of nodes or total number of edges.
3. If a virtual link only connects to components formed by one concept, and this concept it is not itself connected to other concepts, the algorithm may give a higher rank than a page that has actual (not virtual) relations.

Comparing the old and the new method rendered an important difference. The idea presented was to establish if the new method improved the former one. Since there was improvement in some cases, and constraints in other cases, a third method was provided to compare the results. The eigenvector method was modified and adapted to fit a semantic environment. The same principle of backlinks was applied to the ontology and page subgraphs. The results yielded a ranking scheme that was relatively similar to the new method for approximately the first two pages of the results set, and in general the more useful pages for the user.

The simulator compared two algorithms along with a third method for ranking webpages. The limitations of this study were based on the tools used for comparison. To fully evaluate the performance of sematic web environments, an architecture having the same knowledge base, knowledge database, OWL parser, query interface, crawler application, and graphics user interface, would have to be provided for the same ontology and page subgraphs. This study provided results yielded by the algorithms over the same simulator parameters. There are several areas of the simulator that could be further improved, especially from a graphic user interface perspective. A drop-down menu displaying all the concept associations, along with a feedback option after showing the results, could boost the algorithms to progressively display better results with each search like actual search engines today.

The basic idea of the semantic web development was to provide more information between the hyperlinks that have already populated the Web. The information provided was built to bring meaning between the hyperlinks and offer an enhanced method to extract content from websites. In this paper, a similar approach was taken from this idea: the information in the ontologies can be improved and used to provide a more relevant result for the user from a given query. While improvements and constraints were shown in the simulation, the objectives of this study were accomplished by providing an alternative method of ranking.

REFERENCES

[1] Fabrizio, L., A. Sanna, and C. Demartini. "A Realtion-Based Page Rank Algorithm for Semantic Web Search Engines." *IEE Transactions on Knowledge and Data Engineering* 21.1 (2009): 123-36.

[2] Yufe, L., Y. Wang, and X. Huan. "A Relation-Based Search Engine in Semantic Web." *IEEE Transactions on Knowledge and Data Engineering* 19.2 (2007): 273-82.

[3] Kemafor, A., A. Maduko, and A. Sheth. "SemRank: Ranking Complex Relationships Search Results on the Semantic Web." *Proceeding WWW '05 Proceedings of the 14th International Conference on World Wide Web*. 2005. 117-27.

[4] Torsten, P., C. Schalger, and G. Pernul. "A Search Engine for RDF Metadata." *DEXA '04 Proceedings of the Database and Expert Systems Applications, 15th International Workshop*. 2004. 168-72.

[5] Stojanovic, Nenad, Rudi Studer, and Ljiljana Stojanovic. "An Approach for the Ranking of Query Results in the Semantic Web." *Lecture Notes in Computer Science* 2870/2003 (2003): 500-16.

[6] "RDF Concepts." <http://www.w3.org/TR/rdf-concepts/>.

[7] "W3C Semantic Web Activity." <http://www.w3.org/2011/sw/>.

[8] Eyal, Oren, Knud Hinnerk Moller, Simon Scerri, Siegfried Handschuh, and Michael Sintek. *What Are Semantic Annotations?* Science Foundation Ireland under Grants No. SFI/02/CE1/I131 and SFI/04/BR/CS0694 and by the European Commission under the Nepomuk Project FP6-027705.

[9] "W3C Glossaries. Back-link." <http://www.w3.org/2003/glossary/keyword/All/back%20link.html?keywords=back%20link>.

[10] Alvares, J. "The Amazing Librarian." *Plus Magazine*. 1 June 2008. <http://plus.maths.org/content/amazing-librarian>.

[11] Brin, S., and L. Page. "The Anatomy of a Large-Scale Hypertextual Web Search Engine." *Journal Computer Networks and ISDN Systems* April 1 30.1-7 (1998): 107-17.

[12] Lawrence, P., S. Brin, R. Motwani, and T. Winograd. *The PageRank Citation Ranking: Bringing Order to the Web*. Tech. no. SIDL-WP-1999-0120. Standford InfoLab. <http://ilpubs.standford.edu:8090/422/>.

[13] Cohen, S., J. Mamou, Y. Kanza, and Y. Sagiv. "XSearch: A Semantic Search Engine.*" Proc. of Proc. 29th Int'l Conf. Very Large Data Bases*. 2003. 45-56.

[14] Ding, L., T. Finin, A. Joshi, R. Pan, R. S. Cost, Y. Peng, P. Reddivari, V. Doshi, and J. Sachs. "Swoogle: A Search and Metadata Engine for the Semantic Web." *Proc. 13th ACM Int'l Conf. Information and Knowledge Management (CIKM '04)*. 2004. 652-59.

[15] Guha, R., R. McCool, and E. Miller. "Semantic Search." *Proc. 12th Int'l Conf. World Wide Web (WWW '03)*. 2003. 700-09.

[16] Stojanovic, N. "An Explanation-Based Ranking Approach for Ontology-Based Querying." *Proc. 14th Int'l Workshop Database and Expert Systems Applications*. 2003. 167-75.

[17] Cormen, T., C. Leiserson, R. Rivest, and C. Stein. *Introduction to Algorithms*. 3rd ed. The MIT Press, July 31, 1999.